\documentclass[prl,aps,twocolumn,showpacs]{revtex4}
\usepackage{graphicx}
\usepackage{wasysym}
\usepackage{bm}
\usepackage{pstricks}

\newcommand{\beq}{\begin{equation}}
\newcommand{\eeq}{\end{equation}}

\newcommand{\bea}{\begin{eqnarray}}
\newcommand{\eea}{\end{eqnarray}}
\newcommand{\etal}{{\em et al.}}

\def\suhe#1{{\noindent\bf#1:}}
\def\pat#1{{\cal D}#1}

\def\tit#1#2#3#4#5{{#1}{\bf #2}, #3 (#4)}

\def\prep{Phys.\ Rep.\ }
\def\prl{Phys.\ Rev.\ Lett.\ }
\def\pr{Phys.\ Rev.\ }
\def\prb{Phys.\ Rev.\ B\ }

\def\jpco{J.\ Phys.:\ Cond.\ Mat.\ }

\def\jap{J.\ Appl.\ Phys.\ }

\def\natu{Nature\ }

\def\phyb{Physica B\ }

\def\cjp{Can.\ J. Phys.\ }
\def\jacs{J.\ Am.\ Chem.\ Soc.\ }

\def\vecb#1{{\mathbf #1}}
\def\Jm{{\cal J}}
\def\Dm{{\cal D}}
\def\Hm{{\cal H}}
\def\Pm{{\cal P}}

\begin{document}

\title{Why spin ice obeys the ice rules}

\author{S. V. Isakov,$^1$ R. Moessner,$^2$ and S. L. Sondhi$^3$}

\affiliation{$^1$Department of Physics, University of Toronto,
Toronto, Ontario, Canada M5S 1A7} 
\affiliation{$^2$Laboratoire de Physique
Th\'eorique de l'Ecole Normale Sup\'erieure, CNRS-UMR8549, Paris,
France}
\affiliation{$^3$Department of Physics, Princeton University,
Princeton, NJ, USA}


\begin{abstract}
The low temperature entropy of the the spin ice compounds, such as
Ho$_2$Ti$_2$O$_7$ and Dy$_2$Ti$_2$O$_7$, is well described by the
nearest-neighbor antiferromagnetic Ising model on the pyrochlore
lattice, i.e.\ by the ``ice rules''. This is surprising since the
dominant coupling between the spins is their long ranged dipole
interaction. We show that this phenomenon can be understood rather
elegantly: one can construct a model dipole interaction, by adding
terms of shorter range, which yields {\it precisely} the same ground
states, and hence $T=0$ entropy, as the nearest neighbor interaction.
A treatment of the small difference between the model and true dipole
interactions reproduces the numerical work by Gingras \etal\ in
detail. We are also led to a more general concept of projective
equivalence between interactions.
\end{abstract}


\maketitle

\suhe{Introduction}
In 1956, Anderson \cite{andersonice} observed that an Ising
antiferromagnet on the pyrochlore lattice would exhibit a
macroscopic ground state entropy equivalent to that of water
ice \cite{pauling}.  Almost four decades later, in 1997,
such ``spin ice'' behavior was experimentally discovered. 
Harris \etal\ noticed that the pyrochlore compound Ho$_2$Ti$_2$O$_7$,
where only the Holmium ions are magnetic, failed to display any signs
of ordering down to below the ferromagnetic Curie temperature
deduced from its high temperature susceptibility. They proposed
that this was
due to the presence of an easy-axis anisotropy, which leads
to the ferromagnet effectively becoming an Ising pseudospin
antiferromagnet at temperatures well below the anisotropy
strength \cite{harspinice}. This interpretation was strongly 
supported by
a remarkable experiment, in which Ramirez \etal\ showed that the
related compound Dy$_2$Ti$_2$O$_7$ displayed a residual entropy,
${\cal S}_0$, at low temperatures, which was in excellent agreement
with the Pauling estimate for water ice, ${\cal S}_0\approx
(1/2)\ln(3/2)$ \cite{ramspinice}.

However, it was quickly pointed out that the dominant interaction in
these compounds is dipolar due to the large moments on Ho$^{3+}$ and
Dy$^{3+}$, $\mu \approx 10\mu_{B}$, where $\mu_{B}$ is the Bohr
magneton \cite{bangice}. Yet, the dipole interaction is long ranged,
decaying with separation, $r$, as $1/r^3$. Why, then, is the entropy
of the nearest-neighbor model stable to the inclusion of the rest of
this interaction\cite{bangice,watice}?

That this {\it is} the case has been checked in detail by explicit
simulation. Further, two qualitative observations have been made in
the literature.  First, the anisotropic nature of
the dipole interaction precludes any obvious ordering instability
stemming from its long range. It is probably fair to say that it is now
agreed that instead, it leads to a weak ordering instability away from
$\vecb{q}=0$ \cite{bangice,watice,bramwell,dipiceorder},
albeit one  which is
not observed \cite{fn-whatorder}. Second, explicit evaluation of the
Fourier transform of the easy-axis projected dipole interaction to
large (and, using Ewald summation, infinite) distances
yielded a surprise: the
infinite distance result resembled the nearest-neighbor interaction
more than an interaction truncated at, say, 10 or so nearest
neighbors; this `self-screening' (Gingras) indicates that there is 
something special about the dipole interaction 
\cite{dipicetrunc,dipiceewald}.

In this paper we show what it is that is special about spin ice and
dipole interactions. The main insight comes from recent progress in
understanding the correlations dictated by the ice
rule\cite{pyrodip,HKMS3ddimer,hermele03,clhu}: it was shown that the
pyrochlore Ising model dynamically acquires a gauge structure at
$T=0$, which manifests itself in the emergence of dipolar {\em
correlations} as the ice rules are enforced. The crucial observation
is that this gauge structure is (formally, not in origin) exactly the
same as that of `ordinary' magnetostatics; this latter, of course, is
what determines the form of the dipole {\em interactions}.

Most crisply, we use this observation to show that there exists a
slightly modified ``model'' dipole interaction, which differs from
the physical interaction by terms that fall off faster and are small
in magnitude, whose ground states are {\it identically the same} as
those of the nearest neighbor interaction. This accounts for their
identical low temperature entropy.

The mathematical implementation of this insight revolves around
relating two quantities -- dipole interactions, and the operator
enforcing the ice rules -- to the same projector, ${\cal P}$.
Finally, this leads us to a generalizable equivalence between a long
ranged interaction and one of shorter range projected onto its low
energy manifold; this we term ``projective equivalence''.

In the balance of the paper we flesh out these statements. We begin by
recalling the energetics of spin ice. The ground state degeneracy of
the nearest-neighbor antiferromagnet manifests itself in a pair of flat
bands. We show next that a model dipole interaction possessing
the correct long distance form, leads to exactly the same pair of flat
bands and thence to the same ground state manifold, even with
arbitrary  admixture
of the nearest-neighbor interaction (as long as not so overwhelmingly 
antiferromagnetic 
to invert the ordering
of the bands). The deviation of the model
interaction from the true dipole interaction is seen to vanish as an
integrable $r^{-5}$.  Treating this by elementary means we are able to
account for the weak residual dispersion and ordering tendency found
in previous work.  As a byproduct, this analysis makes clear that at
higher temperatures, the correlations arising from the dipole
interaction are not the same as those stemming from the ice rules
alone.  We close with some comments on projective equivalence.

\suhe{Hamiltonian}
The dominant term in the spin ice Hamiltonian is a strong easy easy
axis anisotropy, which allows us to transform the vector spins
$\vecb{S}$ to Ising pseudospins $\sigma=\pm1$ via
$\vecb{S}_{i\alpha}=\sigma_{i\alpha} \vecb{e}_\alpha $ (no sum over
$\alpha$). $\vecb{e}_\alpha$ 
denote the local easy axes of the pyrochlore
lattice, which consists of corner-sharing tetrahedra.  It can be
thought of as a face-centred cubic lattice with a four site basis:
$\vecb{e}_1=(-1,-1,-1)/\sqrt{3}$, $\vecb{e}_2=(1,-1,1)/\sqrt{3}$,
$\vecb{e}_3=(1,1,-1)/\sqrt{3}$, $\vecb{e}_4=(-1,1,1)/\sqrt{3}$; $i$ is
a unit cell index, and $\alpha$ is the sublattice index.

The dipolar spin ice model contains two terms. First, a
nearest-neighbour exchange, $\Jm$, of strength $J$, to which we add a
constant so that its ground states have zero energy; these ground
states obey the ice rules or equivalently the constraint that the
total pseudospin of each tetrahedron vanish. And second, the dipole
interaction $\Dm$, which is summed over {\em all} pairs of sites, with
$D=\mu_0\mu^2/(4\pi r_{\rm nn}^3)$, $\mu$ is the magnetic moment of
the spins, $r_{nn}$ is the nearest-neighbor distance, and
$\vecb{r}_{i\alpha j\beta}$ is the vector separating spins
$\vecb{S}_{i\alpha}$ and $\vecb{S}_{j\beta}$. Defining 
$\Hm\equiv\sum_{i\alpha,j\beta} \Hm_{i\alpha, j\beta} \sigma_{i\alpha}
\sigma_{j\beta}$, the Hamiltonian is:
\bea
  \Hm &=& \sum_{\mathrm{pairs}}
      \sigma_{i\alpha} \left[ J \Jm_{i\alpha, j\beta}
        + D r_{\rm nn}^3\Dm_{i\alpha, j\beta} \right]
    \sigma_{j\beta},
\label{h2}
\\
  \Dm_{i\alpha,j\beta} &=& 
    \frac{\vecb{e}_\alpha \cdot \vecb{e}_\beta}
      {| \vecb{r}_{i\alpha,j\beta} |^3}
    - \frac{3(\vecb{e}_\alpha \cdot \vecb{r}_{i\alpha,j\beta})
         (\vecb{e}_\beta \cdot \vecb{r}_{i\alpha,j\beta})}
      {| \vecb{r}_{i\alpha,j\beta} |^5}
\eea

\suhe{Spectrum and eigenvectors of $\Hm$}
In this section, we establish the connection of the ice rules with the
dipole interaction; this we do by discussing the spectrum and
eigenvectors of the Hamiltonian matrix $\Hm$. These, specifically two
dispersionless bands, will form the basis of our discussion of the
zero temperature entropy and zero and finite temperature correlations
below. 

\begin{figure}[ht]
{
\centerline{\includegraphics[angle=0, width=2.in]{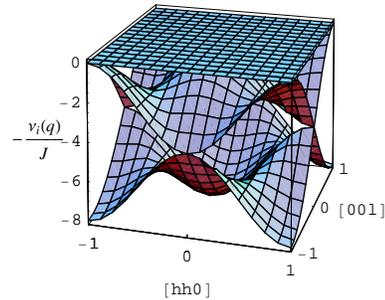}}
\caption{(color online). Mode spectrum of the nearest-neighbor interaction
matrix, $\Jm$,
in the $[hhl]$ plane \cite{canalsgaranin}, $\vecb{q}$ is in units of $2\pi$.
Note that here and on the following plots we plot the eigenvalues, $\nu_i$,
with minus sign.
\label{fig:nn}}
}
\end{figure}

First, consider the spectrum of the adjacency matrix 
$\Jm_{i\alpha,j\beta}$. It is well know that this has a pair
of degenerate flat bands and two dispersive bands (one of them is
gapless at zero wavevector), see Fig.~\ref{fig:nn}. 
The diagonalizing transformation is given in Ref.~\onlinecite{pyrodip}.
We can write the result as the schematic decomposition,
\bea
\Jm = \sum_{\mu=1}^4 \epsilon_\mu(\vecb{q}) 
|v_\mu(\vecb{q}) \rangle \langle
v_\mu(\vecb{q})|
\eea
where for $\mu=1,2$, $ \epsilon_\mu(\vecb{q})\equiv0$. 

\begin{figure}
{
\centerline{\includegraphics[angle=0, width=2.in]{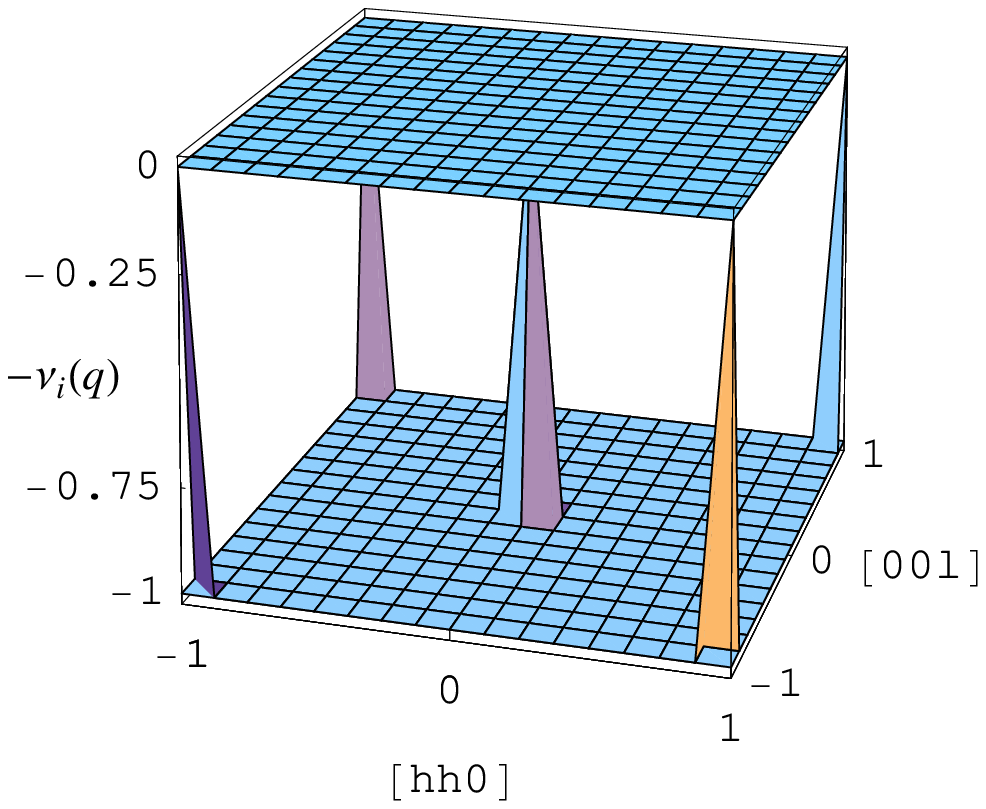}}
\centerline{\includegraphics[angle=0, width=2.in]{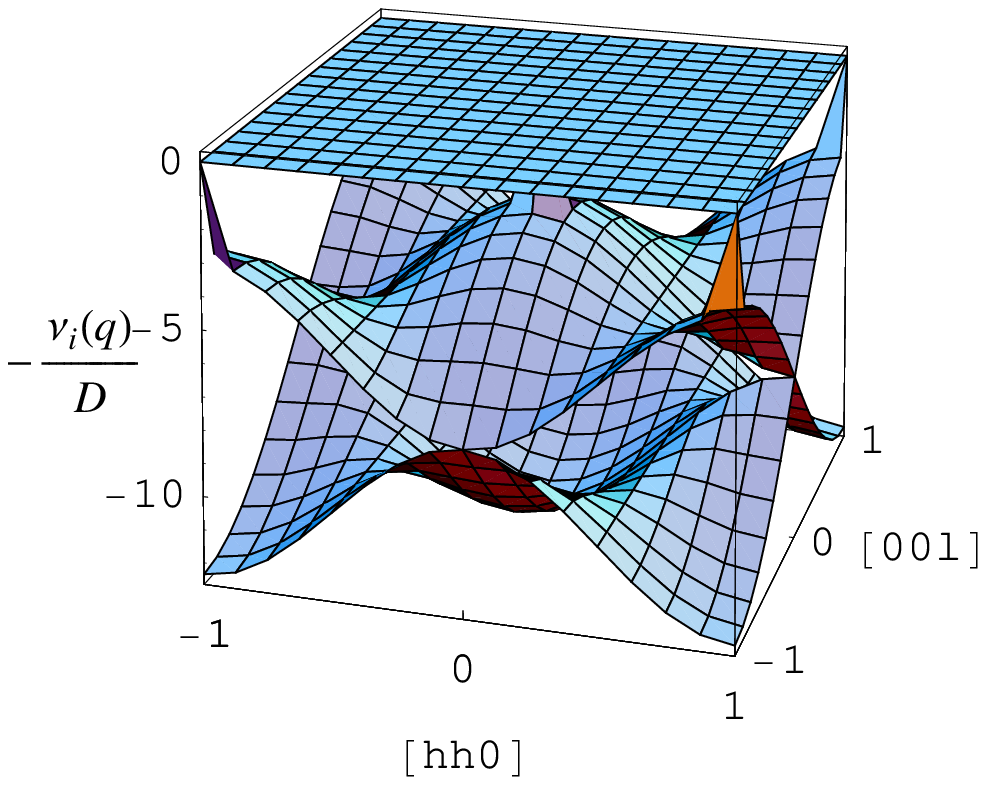}}
\centerline{\includegraphics[angle=0, width=2.in]{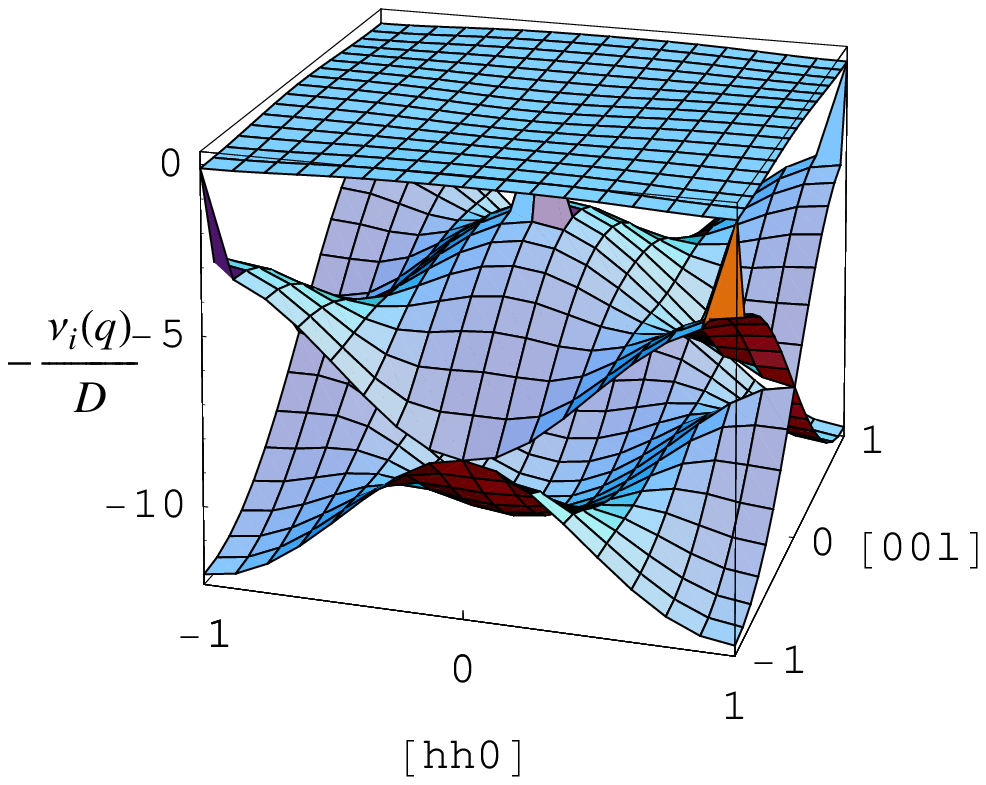}}
\caption{(color online). Mode spectrum of the dipole interaction
(the eigenvalues, $\nu_i$, of the $D r_{\rm nn}^3
\Dm_{ij}^{\alpha\beta}$ matrix).
Top: Model dipole interaction, $\Pm$ only. There are two
degenerate pairs of flat bands. Middle: After inclusion of 
nearest-neighbor correction.  Two bands remain degenerate and
flat. The other two become dispersive.  Adding a nearest-neighbor
superexchange can enhance/suppress this change.  Bottom: Correlation
function plus $\Delta_{ij}^{\alpha\beta}$.  The remaining flat
bands become weakly dispersive.  The middle and bottom plots are
almost identical on this scale. (The spikes in the figure have to with
the special character of the single point at $\vecb{q}=0$ for the
conditionally convergent dipole interaction.)
\label{fig:disp}}
}
\end{figure}

As shown below, at $T=0$, the physics is determined solely by the
modes in the zero-energy flat bands; it will therefore be unchanged for a
family of interaction matrices with any other choice of
$\epsilon_{3,4}>0$ (but keeping the vectors $v$ unchanged). The family
member with $\epsilon_{3,4}\equiv 1$ is then a simple projector
$\Pm=\Pm^2=\sum_{\mu=3}^4 |v_\mu(\vecb{q}) \rangle \langle v_\mu(\vecb{q})|$
(Fig.~\ref{fig:disp}).

This is a longitudinal projector; for formal details, see
Refs.~\onlinecite{pyrodip,HKMS3ddimer,hermele03,clhu}, but in a
nutshell, what happens is this. The ice rules $\sum_{i\in
tet}\sigma_i$ are equivalent to $\nabla\cdot\vecb{S}=0$. Here, the
spins are thought of link variables (`lattice fluxes') on the diamond
lattice, which is dual to the pyrochlore, and $\nabla\cdot$ is the
appropriate lattice divergence. This equation encodes the
statement that longitudinal modes, those with
$\nabla\cdot\vecb{S}\neq0$, cost energy, whereas transverse ones do
not. The energetically enforced constraint $\nabla\cdot\vecb{S}=0$ is
at
the origin of the `emergent gauge structure' of spin ice, as it can be
resolved by transforming to a gauge field (vector potential), 
$\vecb{S}=\nabla\times\vecb{A}$. 

What are the matrix elements of $\Pm$ in real space? Working backwards
from the form of the spin ice correlations obtained in 
\cite{pyrodip,HKMS3ddimer,hermele03,clhu},
one can read off that, asymptotically at large distances,
$\Dm\propto\Pm$. 
More precisly, defining a correction
term $\Delta$ through 
\beq
  \Dm_{i\alpha,j\beta}=\frac{8\pi}{3} \Pm_{i\alpha,j\beta}
    + \Delta_{i\alpha,j\beta} \ ,
\label{d1}
\eeq
analyticity and symmetry considerations give 
$\Delta_{i\alpha,j\beta}\sim O(r_{i\alpha
j\beta}^{-5})$.

Thus, an interaction $\Pm$ (i) has the same long-distance form as $\Dm$ and
(ii) has the same eigenvectors and same ground-state manifold as $\Jm$. With
this in hand, we can now understand all the important qualitative features of
the spectrum of $\Hm$ analytically and gain a quantitative understanding with
computations that do not require us explicitly to treat the conditional
convergence of dipole sums. (Note that we again add an overall constant 
to the energy so that the flat bands occur at zero energy).

Starting with the model dipole interaction $\Pm$
(Fig.~\ref{fig:disp}, top panel), it is trivial to add in the
superexchange as $\Jm$ and $\Pm$ have the same
eigenvectors. The same is true of the nearest-neighbor pieces of
$\Delta_{i\alpha,j\beta}$, by far its largest matrix elements, which
are also proportional to $\Jm$.  The net result is that the lower pair
of flat bands remain flat while the upper pair acquire the same
dispersion as $\Jm$---this is illustrated in the middle panel of
Fig.~\ref{fig:disp}.

Thus far we have shown that ``much'' of $\Hm$ is characterized by
a pair of low lying flat bands. What remain, the matrix elements
of $\Delta_{i\alpha,j\beta}$ {\it beyond the nearest-neighbor 
distance}, are small. Thus their inclusion does
not modify the gross features significantly, preserving a spectrum
similar to that of the nearest-neighbor model. Indeed, adding
$\Delta_{ij}^{\alpha\beta}$ beyond the nearest-neighbor distance
weakly splits the remaining pair of flat bands, which acquire a small
dispersion; the other pair of bands is barely modified on the scale of
their dispersion; this is shown in the figure's bottom panel.

The full $\Delta$ was included by adding its numerical Fourier
transform to that of $\Jm$ and $\Pm$ obtained analytically, and
diagonalising the resulting $4\times4$ matrix. 
With $\Delta_{i\alpha,j\beta}$ small and
decaying fast, its Fourier transform is quickly and absolutely
convergent; we have checked that the results are essentially
independent of the truncation distance for $r_c\ge12r_{nn}$. We would
like to stress that this truncation is not equivalent to the
truncation of the long-range dipole interaction since in our case the
main long distance part of the dipole interaction is already
contained in $\Pm$.

\begin{figure}[ht]
{
\centerline{\includegraphics[angle=0, width=2.2in]{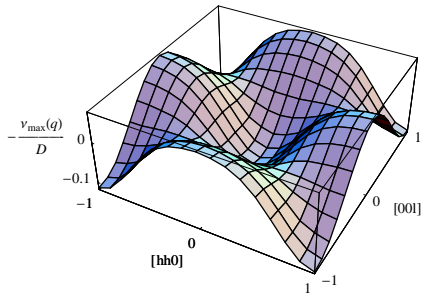}}
\caption{(color online).
Maximum eigenvalue, $\nu_{\rm max}$, of the dipole matrix in the $[hhl]$
plane, $\vecb{q}$ is in units of $2\pi$, $r_c=16r_{\rm nn}$. 
\label{fig:minval}}
}
\end{figure}

In Fig.~\ref{fig:minval}, we show the band of maximal eigenvalues of
the dipole matrix. It is weakly dispersive because
$\Delta_{ij}^{\alpha\beta}$ is small. Our results are in quantitative
agreement with the Ewald summation results \cite{dipiceewald}.

\suhe{Ground states and entropy} 
The nearest neighbor interaction by
itself gives rise to a macroscopic entropy at $T=0$. We now show that
any combination of the nearest neighbour, $\Jm$ and the model dipole, $\Pm$,
interactions
leads to {\it precisely} the same ground states, and hence
entropy.

First, note that the ground states of $\Jm$ have zero energy; 
an arbitrary ground state,
$|0\rangle$ can thus be written as a linear combination of modes in
the flat zero-energy bands only, as an admixture of other bands would 
lead to a non-zero energy:
\bea
|0\rangle=\sum_{\bf q}\sum_{\mu=1}^2a_\mu({\bf q})|v_\mu({\bf q})\rangle\ .
\eea 
The hard-spin condition imposes constraints on the
amplitudes $a_\mu({\bf q})$, which we do not need to 
resolve explicitly.

It suffices to note that, 
having written the ground states of $\Jm$ in terms of the flat bands,
it follows that they remain ground states upon admixing the model
dipole interaction, $\Pm$, as this only affects the relative position
of the excluded positive
energy bands $\mu=3,4$. Indeed, this implies that the set of ground 
states of $\Jm$ and $\Pm$ are identical.

The addition of $\Delta$
will lift this degeneracy but, as we saw, only 
weakly. Consequently the low temperature
entropy of spin ice will be quite close to that of ice before it goes
away at the lowest temperatures as the system orders\cite{fn-whatorder}.

\suhe{Correlations}
This explains why the similarity of the spectrum 
is sufficient to yield the correct low-temperature
physics: for the model dipole interaction, the ground state correlations
are exactly those averaged over the ice rule manifold. 
However, this equivalence between the
model dipole and the nearest neighbor antiferromagnetic Ising
model breaks down at nonzero temperatures.
For the nearest-neighbor Ising model, the presence of thermally
activated
ice rule violating defects leads to an exponential decay 
of correlations on a length scale diverging as $\xi\sim\exp(2J/3T)$ 
at low $T$.

By contrast, for the dipole problem
the long range of the interaction implies long ranged correlations
at {\it any} temperature.
This is already evidenced by the first term in the
high-temperature expansion of
\bea
\langle\vecb{S}_{i\alpha}\cdot\vecb{S}_{j,\beta}\rangle
\propto-\Hm_{i\alpha,j\beta}/T\sim O(r_{i\alpha,j\beta}^{-3})\ .
\eea
In fact, in a
saddle-point treatment for $\Pm$, one can show that this holds for any
temperature, and to all orders in corrections to the saddle point.  

{\bf Projective equivalence:} The mathematics underlying our analysis
of dipolar spin ice can be generalized. One can construct other
exchange matrices $\Jm^\prime$ which share their low-lying flat bands
(and its eigenvectors) with interactions $\Dm^\prime$ of longer range;
we should note though, that generically, neither the $\Jm^\prime$ nor
the $\Dm^\prime$ are of bounded range. In this fashion, we find
pairs of interactions which are equivalent under projection to the
flat bands; this we term ``projectively equivalent''.  

The miracle of spin ice is hence twofold: first, that the physical
dipole interaction restricted to the site dependent easy axes on the
pyrochlore lattice provides one member of such a pair; and second,
thanks to its emergent gauge structure, the other member of the pair
is the classic ice problem dating back to Bernal, Fowler and Pauling.
In short: dipolar spins are ice because ice is dipolar.

\suhe{Acknowledgements}
We would like to thank Matt Enjalran and Michel Gingras for generously
sharing their partially published Ewald summation results with us. We
are also grateful to David Huse, Karol Gregor and Werner Krauth for
collaboration on closely related work. We thank the
above, Chris Henley, and Oleg Tchernyshyov for useful
discussions.  This work was in part supported by the Minist\`ere de la
Recherche et des Nouvelles Technologies, by the NSF 
[PHY99-07949 (at KITP), 0213706] and by the David and Lucile Packard
Foundation.

\end{document}